# Exponential Lower Bounds For Policy Iteration

John Fearnley

Department of Computer Science, University of Warwick, UK
john@dcs.warwick.ac.uk

**Abstract.** We study policy iteration for infinite-horizon Markov decision processes. It has recently been shown policy iteration style algorithms have exponential lower bounds in a two player game setting. We extend these lower bounds to Markov decision processes with the total reward and average-reward optimality criteria.

## 1 Introduction

The problem of finding an optimal policy for infinite-horizon Markov decision process has been widely studied [8]. Policy iteration is one method that has been developed for this task [5]. This algorithm begins by choosing an arbitrary policy, and then iteratively improves that policy by modifying the policy so that it uses different actions. For each policy, the algorithm computes a set of actions that are switchable, and it then chooses some subset of these actions to be switched. The resulting policy is guaranteed to be an improvement.

The choice of which subset of switchable actions to switch in each iteration is left up to the user: different variants of policy iteration can be created by giving different rules that pick the subset. Traditionally, policy iteration algorithms use a greedy rule that switches every state with a switchable action. Greedy policy iteration will be the focus of this paper.

Policy iteration has been found to work well in practice, where it is used as an alternative to linear programming. Linear programming is known to solve the problem in polynomial time. However, relatively little is known about the complexity of policy iteration. Since each iteration yields a strictly improved policy, the algorithm can never consider the same policy twice. This leads to a natural exponential bound on the number of iterations before the algorithm arrives at the optimal policy. The best upper bounds have been provided by Mansour and Singh [6], who showed that greedy policy iteration will terminate in $O(k^n/n)$ iterations, where $k$ is the maximum number of outgoing actions from a state.

Melekopoglou and Condon have shown exponential lower bounds for some simple variants of policy iteration [7]. The policy iteration algorithms that they consider switch only a single action in each iteration. They give a family of examples upon which these policy iteration algorithms take $2^n - 1$ steps. It has been a long standing open problem as to whether exponential lower bounds could be shown for greedy policy iteration. The best lower bound that has been shown so far is $n + 6$ iterations [2].

Policy iteration is closely related to the technique of strategy improvement for two player games. Friedmann [4] has recently found a parity game which forces which forces the strategy improvement algorithm of Vöge and Jurdziński [9] to take an exponential number of steps. It has been shown that this example can be used to show exponential lower bounds for strategy improvement algorithms for other prominent types of two player game [1].

*Our contribution.* Friedmann's example relies on the fact that there are two players in a parity game. We show how Friedmann's example can be adapted to provide exponential lower bounds for policy iteration on Markov decision processes. We present an example that provides an exponential lower bound for the total reward criterion, and we also argue that the same example provides an exponential lower bound for the average-reward criterion.

## 2 Preliminaries

A Markov decision process consists of a set of states $S$, where each state $s \in S$ has an associated set of actions $A_s$. For a given state $s \in S$ and action $a \in A_s$ the function $\mathrm{r}(s, a)$ gives the reward for choosing the action $a$ in the state $s$. Given two states $s$ and $s'$, and an action $a \in A_s$, the function $\mathrm{p}(s'|s, a)$ gives the probability of moving to state $s'$ when the action $a$ is chosen in state $s$. This is a probability distribution, so $\sum_{s' \in S} \mathrm{p}(s'|s, a) = 1$ for all $s$ and $a \in A_s$.

A policy $\pi : S \to A_s$ is a function that selects one action at each state. For a given starting state $s_0$, a run that is consistent with a policy $\pi$ is an infinite sequence of states $\langle s_0, s_1, \dots \rangle$ such that $p(s_{i+1}|s_i, \pi(s_i)) > 0$ for all $i$. The set $\Omega^\pi_{s_0}$ contains every consistent run from $s_0$ when $\pi$ is used. A probability space can be defined over these runs using the $\sigma$-algebra that is generated by the cylinder sets of finite paths starting at $s_0$. The cylinder set of a finite path contains every infinite path that has the finite path as a prefix. If we fix the probability of the cylinder set of a finite path $\langle s_0, s_1, s_2, \dots s_k \rangle$ to be $\prod_{i=0}^{k-1}(s_{i+1}|s_i, \pi(s_i))$, then standard techniques from probability theory imply that there is a unique extension to a probability measure $\mathbb{P}^\pi_{s_0}(\cdot)$ on the $\sigma$-algebra [3]. Given a function that assigns a value to each consistent run $f : \Omega \to \mathbb{R}$, we define $\mathbb{E}^\pi_{s_0}\{f\}$ to be the expectation of this function in the probability space.

The value of a state $s$ when a policy $\pi$ is used varies according to the choice of optimality criterion. In the total reward criterion the value is $\mathrm{Val}^\pi(s) = \mathbb{E}^\pi_s\{\sum_{i=0}^{\infty} \mathrm{r}(s_i, s_{i+1})\}$, and for the average-reward criterion the value is $\mathrm{Val}^\pi_\mathcal{A}(s) = \mathbb{E}^\pi_s\{\liminf_{N \to \infty} \frac{1}{N} \sum_{i=0}^{N} \mathrm{r}(s_i, s_{i+1})\}$. The computational objective is to find the optimal policy $\pi^*$, which is the policy that maximizes the value function for every starting state. We define the value of a state to be the value of that state when an optimal policy is being used. That is, we define $\mathrm{Val}(s) = \mathrm{Val}^{\pi^*}(s)$ and $\mathrm{Val}_\mathcal{A}(s) = \mathrm{Val}^{\pi^*}_\mathcal{A}(s)$ for every state $s$.

For each optimality criterion, it has been shown that the value of each state can be characterised by the solution of a system of optimality equations [8]. For

the total reward criterion these optimality equations are, for every state $s$:

$$V(s) = \max_{a \in A_s}(\text{r}(s,a) + \sum_{s' \in S} p(s'|s,a) \cdot V(s')) \quad (1)$$

For the average-reward criterion we have two types of optimality equation, which must be solved simultaneously. The first of these are called the gain equations:

$$G(s) = \max_{a \in A_s}(\sum_{s' \in S} p(s'|s,a) \cdot G(s')) \quad (2)$$

Secondly we have the bias equations. If $M_s = \{a \in A_s \ : \ G(s) = \sum_{s' \in S} p(s'|s,a) \cdot G(s')\}$ is the set of actions that satisfy the gain equation at the state $s$, then the bias equations are defined as:

$$B(s) = \max_{a \in M_s}(\text{r}(s,a) - G(s) + \sum_{s' \in S} p(s'|s,a) \cdot B(s')) \quad (3)$$

We have that solutions to these equations characterise the value of every state. That is, we have $\text{Val}(s) = V(s)$ and $\text{Val}_\mathcal{A}(s) = G(s)$, for every state $s$. We can also obtain an optimal policy by setting $\pi^*(s) = a$, where $a$ is an action that achieves the maximum in the optimality equation.

## 3 Policy Iteration

Policy iteration is a method for solving the optimality equations that we presented in Section 2. We will begin by describing policy iteration for the total reward criterion. For every policy $\pi$ that the algorithm considers, it will compute the value $\text{Val}^\pi(s)$ of the policy at every state $s$, and check whether this is a solution of the optimality equation (1). The value of the policy can be obtained by computing the solution to:

$$\text{Val}^\pi(s) = \text{r}(s,a) + \sum_{s' \in S} p(s'|s,a) \cdot \text{Val}^\pi(s') \quad (4)$$

If the value of $\pi$ satisfies the optimality equation (1) at every state, then a solution has been found and the algorithm can terminate. Otherwise, we define the appeal for each action $a \in A_s$ in the policy $\pi$ to be: $\text{Appeal}^\pi(s,a) = \text{r}(s,a) + \sum_{s' \in S} p(s'|s,a) \cdot \text{Val}^\pi(s')$. If the policy $\pi$ does not satisfy the optimality equation then there must be at least one action $a$ at a state $s$ such that $\text{Appeal}^\pi(s,a) > \text{Val}^\pi(s)$. We say that an action with this property is switchable in $\pi$. Switching an action $a \in A_t$ in a policy $\pi$ creates a new policy $\pi'$ where $\pi'(s) = a$ if $s = t$, and $\pi'(s) = \pi(s)$ for every other state $s$. The set of switchable actions is important because it can be shown that switching any subset of switchable actions will create an improved policy.

**Theorem 1 ([8]).** *If $\pi$ is a policy and $\pi'$ is a policy that is obtain by switching some subset of switchable actions in $\pi$ then $\text{Val}^{\pi'}(s) \geq \text{Val}^\pi(s)$ for every state $s$, and there is some state in which the inequality is strict.*

Policy iteration begins by choosing an arbitrary policy for the MDP. In every iteration it computes the set of switchable actions, and then picks some subset of these actions to switch in the current policy. This creates a new policy which will be considered in the subsequent iteration. Since policy iteration only ever switches switchable actions, Theorem 1 implies that it cannot visit the same policy twice. This is because repeating a policy would require the value of some state to decrease. Since there are a finite number of policies, the algorithm must eventually arrive at a policy with no switchable actions. This policy clearly satisfies the optimality equation (1), and policy iteration can terminate.

Note that any subset of of switchable actions can be chosen in each iteration of the algorithm, and the choice of subset affects the behaviour of the algorithm. In this paper we study the greedy policy iteration algorithm, which selects the most appealing switchable action at every state. For every state $s$ where equation (1) is not satisfied, the algorithm will switch the action: $\mathrm{argmax}_{a \in A_s}(\mathrm{Appeal}^\pi(s, a))$.

Policy iteration for the average-reward criterion follows the pattern, but it uses uses optimality equations (2) and (3) to decide which actions are switchable in a given policy. For each policy it computes a solution to:

$$G^\pi(s) = \sum_{s' \in S} p(s'|s, \pi(s)) \cdot G^\pi(s)$$

$$B^\pi(s) = \mathrm{r}(s, \pi(s)) - G^\pi(s) + \sum_{s' \in S} p(s'|s, \pi(s)) \cdot B^\pi(s')$$

An action $a \in A_s$ is switchable if either $\sum_{s' \in S} p(s'|s, a) \cdot G^\pi(s') > G^\pi(s)$ or if $\sum_{s' \in S} p(s'|s, a) \cdot G^\pi(s') = G^\pi(s)$ and: $\mathrm{r}(s, a) - G^\pi(s) + \sum_{s' \in S} p(s'|s, a) > B^\pi(s)$.

## 4 Exponential Lower Bounds For The Total Reward Criterion

In this section we will describe a family of examples that force policy iteration for the total reward criterion to take an exponential number of steps. Due to the size and complexity of the example, we will break the example down into several component parts, which will be presented separately.

The example will actually contain very few actions that are probabilistic. An action $a \in A_s$ is deterministic if there is some state $s'$ such that $\mathrm{p}(s'|s, a) = 1$. For the sake of convenience, we will denote actions of this form as $(s, s')$. We also overload our previous notations: the notation $\pi(s) = s'$ indicates that $\pi$ chooses the deterministic action from $s$ to $s'$, the function $\mathrm{r}(s, s')$ gives the reward of this action, and $\mathrm{Appeal}^\pi(s, s')$ gives the appeal of this action under the policy $\pi$.

Since we are working with the total reward criterion, care must be taken to ensure that the value of a policy remains well defined. For this purpose, the example will contain a sink state $c_{n+1}$ that has a single action $(c_{n+1}, c_{n+1})$ with reward 0. This will be an absorbing state, in the sense that every run of the MDP from every starting state will eventually arrive at the state $c_{n+1}$, for every policy that is considered by policy iteration. This will ensure that the value of each state remains finite throughout the execution of the algorithm.

We will give several diagrams for parts of the example, such as the diagram given in Figure 1. States are represented by boxes, and the name of a state is printed on the box. Actions are represented by arrows: deterministic actions are represented as an arrow from one state to another, and probabilistic actions are represented as arrows that split, and end at multiple states. The probability distribution is marked after the arrow has split, and the reward of the action is marked before the arrow has split.

Our overall goal is to construct an example that forces policy iteration to mimic the behaviour of a binary counter. Each policy will be associated with some configuration of a binary counter, and the exponential lower bound will be established by forcing policy iteration to pass through at least one policy for every possible configuration of the binary counter. If the bits of this counter are indexed 1 through $n$, then there are two conditions that are sufficient enforce this behaviour. Firstly, a bit with index $i$ should become 1 only after all bits with index $j < i$ are 1. Secondly, when the bit with index $i$ becomes 1, every bit with index $j < i$ must be set to 0. Our exposition will be follow this structure: in section 4.1 we will describe how each policy corresponds to a configuration of a binary counter, in section 4.2 we will show how the first condition is enforced, and in section 4.3 we will show how the second condition is enforced.

### 4.1 A Bit

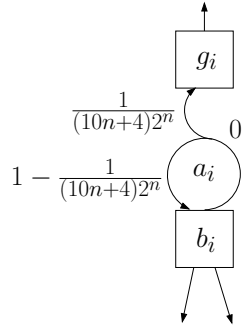

**Fig. 1.** The structure for the bit with index $i$.

The example will contain $n$ instances of structure shown in Figure 1, which will represent the bits of a binary counter. We will represent the configuration of a binary counter as a set $B \subseteq \{1, 2, \ldots n\}$ that contains the indices of the bits that are 1. A policy $\pi$ represents a configuration $B$ if $\pi(b_i) = a_i$ for every index $i \in B$, and $\pi(b_i) \neq a_i$ for every every index $i \notin B$. For a set of natural numbers $B$ we define $B^{>i}$ to be the set $B \setminus \{k \in \mathbb{N} \ : \ k \leq i\}$. We define analogous notations for the relations $<$, $\geq$, and $\leq$.

The actions $a_i$ will be the only probabilistic actions in the example. It is worth noting that when the action $a_i$ is chosen at $b_i$ the effect, under the total reward criterion, is identical to a deterministic action $(b_i, g_i)$ with reward 0. The fact that it takes an expected $(10n + 4)2^n$ steps to move from $b_i$ to the $g_i$ using the action $a_i$ is irrelevant because the reward of $a_i$ is 0, and these steps will have no effect on the total reward.

**Proposition 2.** *For every policy $\pi$, if $\pi(b_i) = a_i$ then $\mathrm{Val}^\pi(b_i) = \mathrm{Val}^\pi(g_i)$.*

The reason why the given probabilities have been chosen for the action $a_i$ is that the value of the state $g_i$ will never exceed $(10n + 4)2^n$.

**Assumption 3** *For every policy $\pi$ we have $\mathrm{Val}^\pi(b_i) > 0$ and $\mathrm{Val}^\pi(g_i) \leq (10n + 4)2^n$.*

Although the action $a_i$ behaves like a deterministic action when it is chosen at $b_i$, it behaves differently when it is not chosen. A deterministic action $(b_i, g_i)$ would have $\mathrm{Appeal}^\pi(b_i, g_i) = \mathrm{Val}^\pi(g_i)$ in every policy. By contrast, when $a_i$ is not chosen by a policy $\pi$, we can show that the appeal of $a_i$ is at most $\mathrm{Val}^\pi(b_i)+1$.

**Proposition 4.** *Suppose that Assumption 3 holds. If $\pi$ is a policy such that $\pi(b_i) \neq a_i$ then $\mathrm{Appeal}^\pi(b_i, a_i) < \mathrm{Val}^\pi(b_i) + 1$.*

This is the key property that will allow us to implement a binary counter. The value of the state $g_i$ could be much larger than the value of $b_i$. However, we are able to prevent policy iteration from switching the action $a_i$ by ensuring that there is always some other action $x$ such that $\mathrm{Appeal}(b_i, x) \geq \mathrm{Val}^\pi(b_i) + 1$.

### 4.2 Switching the Smallest Bit

In this section we will give a full description of the example, and we will show how policy iteration can only switch the state $b_i$ to $a_i$ after every state $b_j$ with $j < i$ has been switched to $b_j$.

Figure 2 shows one of the key structures in the example, which is called the deceleration lane. In the previous section we argued that an action $(b_i, x)$ with $\mathrm{Appeal}^\pi(b_i, x) \geq \mathrm{Val}^\pi(b_i) + 1$ is required in every policy $\pi$ with $\pi(b_i) \neq a_i$ to prevent policy iteration from switching the action $a_i$. The deceleration is the structure that ensures these actions will exist.

The states $x$ and $y$ both have outgoing actions that will be specified later. For now, we can reason about the behaviour of the deceleration lane by assuming that the value of $y$ is larger than the value of $x$.

**Assumption 5** *For every policy $\pi$ we have $\mathrm{Val}^\pi(y) > \mathrm{Val}^\pi(x)$.*

The initial policy for the deceleration lane is the one in which every state $d_k$ chooses the action $(d_k, y)$. It is not difficult to see that the only switchable action in this policy is $(d_1, d_0)$. This is a general trend: the action $(d_j, d_{j-1})$ can only be switched after every action $(d_k, d_{k-1})$ with $1 \leq k < j$ has been switched.

**Fig. 2.** The deceleration lane.

Therefore, policy iteration will take $2n$ steps to arrive at the optimal policy for the deceleration lane. Formally, for every $j$ in the range $0 \leq j \leq 2n$ we define:

$$\pi_j(s) = \begin{cases} d_{k-1} & \text{if } s = d_k \text{ and } 1 \leq k \leq j, \\ y & \text{otherwise.} \end{cases} \quad (5)$$

**Proposition 6.** *Suppose that Assumption 5 holds. Applying policy iteration to $\pi_0$ produces the sequence of policies $\langle \pi_0, \pi_1, \ldots, \pi_{2n} \rangle$.*

**Fig. 3.** The outgoing actions from the state $b_i$.

Figure 3 shows how each state $b_i$ is connected to the deceleration lane. Of course, since we have not yet specified the outgoing actions from the states $f_i$, we cannot reason about their appeal. These actions will be used later to force the state $b_i$ to switch away from the action $a_i$ as the binary counter moves between configurations. For now, we can assume that these actions are not switchable.

**Assumption 7** *We have $\text{Appeal}^\pi(b_i, f_j) < \text{Val}^\pi(b_i)$ for every policy $\pi$ and every action $(b_i, f_j)$.*

We now describe the behaviour of policy iteration for every index $i \notin B$. The initial policy for the state $b_i$ will choose the action $(b_i, y)$. In the first iteration the action $(b_i, d_{2i})$ will be switched, but after this the action chosen at $b_i$ follows the deceleration lane: policy iteration will switch the action $(b_i, d_k)$ in the iteration immediately after it switches the action $(d_k, d_{k-1})$. Since $r(b_i, d_k) + r(d_k, d_{k-1}) = r(b_i, d_{k-1}) + 1$, this satisfies the condition that prevents the action $a_i$ being switched at $b_i$. Formally, for every $j$ in the range $0 \leq j \leq 2i+1$ we define:

$$\pi_j^o(s) = \begin{cases} \pi_j(s) & \text{if } s = d_j \text{ for some } j, \\ y & \text{if } j = 0 \text{ and } s = b_i \\ d_{2i} & \text{if } j = 1 \text{ and } s = b_i \\ d_{j-1} & \text{if } 2 \leq j \leq 2i+1 \text{ and } s = b_i. \end{cases}$$

**Proposition 8.** *Suppose that Assumptions 3, 5, and 7 hold. When policy iteration is applied to $\pi_0^o$ it will will produce $\langle \pi_0^o, \pi_1^o, \ldots \pi_{2i+1}^o \rangle$.*

We can now see why a bit with index $i$ can only be set to 1 after all bits with index $j$ such that $j < i$ have been set to 1. Since each state $b_i$ has $2i$ outgoing actions to the deceleration lane, policy iteration is prevented from switching the action $a_i$ for $2i + 2$ iterations. Therefore, policy iteration can switch $a_i$ at the state $b_i$ at least two iterations before it can switch $a_j$ at a state $b_j$ with $j > i$.

The second important property of the deceleration lane is that it can be reset. If at any point policy iteration arrives at a policy $\pi_j^o$ in which $\text{Val}^{\pi_j^o}(x) > \text{Val}^{\pi_j^o}(y) + 6n + 1$ then policy iteration will switch the actions $(d_k, x)$ for all $k$ and the action $(b_i, x)$ for every $i \in B$, to create a policy $\pi'$. The reason why these actions must be switched is that the largest value that a state $d_k$ or $b_i$ can obtain in a policy $\pi_j^o$ is $\text{Val}^{\pi_j^o}(y) + 6n + 1$. Now suppose that $\text{Val}^{\pi'}(y) > \text{Val}^{\pi'}(x) + 4n$. If this is true, then policy iteration will switch the actions $(d_k, y)$ and the action $(b_i, y)$, and it therefore arrives at the policy $\pi_0^o$. The ability to force the deceleration lane to reset by manipulating the difference between the values of $y$ and $x$ will be used in the next section.

We now turn our attention to the states $b_i$ where $i \in B$. Policy iteration should never switch away from the action $a_i$ at these states irrespective of the state of the deceleration lane. Since we have not yet specified the outgoing actions of $g_i$, we need to assume that the value of $b_i$ is large enough to prevent the actions $(b_i, d_k)$ being switchable.

**Assumption 9** *For every policy $\pi$, if $i \in B$ then $\text{Val}^\pi(b_i) > \text{Val}^\pi(y) + 6n + 1$.*

With this assumption holds, the state $b_i$ will not be switched away from the action $a_i$. Formally, for $j$ in the range $2 \leq j \leq 2i$ we define:

$$\pi_j^c(s) = \begin{cases} \pi_j(s) & \text{if } s = d_j \text{ for some } j, \\ a_i & \text{if } s = b_i. \end{cases}$$

**Proposition 10.** *Suppose that Assumptions 5, 7, and 9 hold. When policy iteration is applied to $\pi_0^c$ it will produce the sequence $\langle \pi_0^c, \pi_1^c, \ldots \pi_{2i}^c \rangle$.*

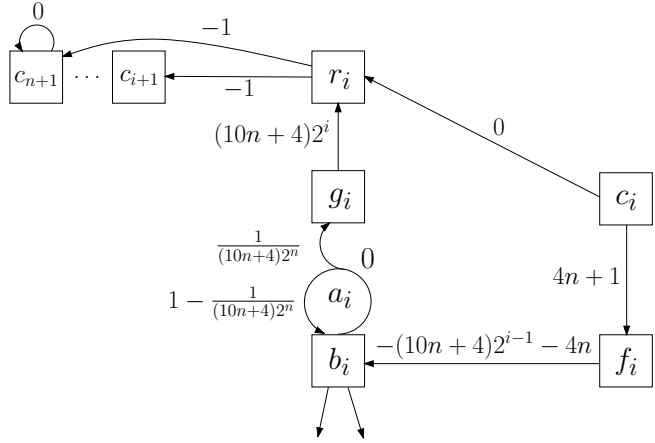

**Fig. 4.** The structure associated with the state $b_i$.

Figure 4 shows the structure that is associated with each state $b_i$. We complete the example by specifying the outgoing actions from $x$ and $y$: there is an action $(y, c_i)$ with reward 0 for every $i$ in the range $1 \le i \le n+1$, there is an action $(x, f_i)$ with reward 0 for every $i$ in the range $1 \le i \le n$, and there is an action $(x, c_{n+1})$ with reward $-1$.

The idea is that the state $c_i$ should use the action $(c_i, f_i)$ only when the index $i$ is a member of $B$. Moreover, the state $r_i$ should use the action $(c_i, r_j)$ where $j \in B$ is the smallest bit that is both larger than $i$ and a member of $B$. The state $x$ should use the action $(x, f_j)$ and the state $y$ should use the action $(y, c_j)$ where $j$ is the smallest index that is a member of $B$.

Formally, for each configuration $B$ we define a policy $\pi^B$ for these states. We define $\pi^B(c_i) = (c_i, f_i)$ if $i \in B$ and $\pi^B(c_i) = (c_i, r_i)$ if $i \notin B$. We define $\pi^B(r_i) = (r_i, c_j)$ where $j = \min(B^{>i} \cup \{n+1\})$. We define $\pi^B(y) = (y, c_j)$ where $j = \min(B \cup \{n+1\})$. We define $\pi^B(x) = (x, f_j)$ where $j = \min(B)$ if $B \ne \emptyset$, and we define $\pi^B(x) = (x, c_{n+1})$ if $B = \emptyset$.

We can now define the sequence of policies that policy iteration will pass through for each configuration $B$. This definition combines the partial policies $\pi_j$, $\pi_j^o$, $\pi_j^c$, and $\pi^B$ to give a complete policy $\pi_j^B$. If $i = \min(\{i \notin B : 1 \le i \le n\})$ then we define $\text{Sequence}(B) = \langle \pi_1^B, \pi_2^B, \ldots \pi_{2i+1}^B \rangle$, where:

$$\pi_j^B(s) = \begin{cases} \pi_j(s) & \text{if } s = d_k \text{ for some } k, \\ \pi_j^c(s) & \text{if } s = b_j \text{ where } j \in B, \\ \pi_j^o(s) & \text{if } s = b_j \text{ where } j \notin B. \\ \pi^B(s) & \text{otherwise.} \end{cases}$$

We can now see why the assumptions that we have made are true in the full example. For example, in Assumption 3 we asserted that $\text{Val}^\pi(g_i) \le (10n+4)2^n$.

This holds for every policy $\pi_j^B$ because by following this policy from the state $g_i$ we pass through $r_i$ followed by $c_j, f_j, b_j, g_j$, and $r_j$ for every index $j \in B^{>i}$, before arriving at the sink state $c_{n+1}$. Therefore, the value of the state $g_i$ under the policy $\pi_j^B$ can be at most $\sum_{l=i+1}^{n}(10n+4)(2^l - 2^{l-1}) + (10n+4)2^i = (10n+4)2^n$. The other assumptions that we have made can be also be shown to be true for every policy $\pi_j^B$.

**Proposition 11.** *For every configuration $B$ we have that Assumptions 3, 5, 7, and 9 hold for every policy $\pi$ in $\text{Sequence}(B)$.*

Our previous propositions have done most of the work in showing that if policy iteration is applied $\pi_0^B$, then it will pass through $\text{Sequence}(B)$. To complete the proof it is sufficient to note that policy iteration never switches away from the policy $\pi^B$ at the states $c_i, r_i, x$, and $y$.

**Proposition 12.** *When policy iteration is applied to $\pi_0^B$ policy iteration will pass through the sequence of policies given by $\text{Sequence}(B)$.*

### 4.3 Moving Between Configurations

In this section we will describe the behaviour of policy iteration after the final policy in $\text{Sequence}(B)$ has been reached. Throughout this section we define $i = \min(\{j \notin B : 1 \le j \le n\})$ to be the smallest index that is not in the configuration $B$, and we define $B' = B \cup \{i\} \setminus \{1, 2, \ldots i-1\}$. Our goal is to show that policy iteration moves from the policy $\pi_{2i+1}^B$ to the policy $\pi_0^{B'}$.

The first policy that policy iteration will move to is identical to the policy $\pi_{2i+2}^B$, with the exception that the state $b_i$ is switched to the action $a_i$. We define:

$$\pi_{R1}^B(s) = \begin{cases} a_i & \text{if } s = b_i, \\ \pi_{2i+2}^B(s) & \text{otherwise.} \end{cases}$$

This occurs because the state $b_i$ only has $2i$ actions of the form $(b_i, d_k)$. Therefore, once the policy $\pi_{2i+1}^B$ is reached there will no be no action of the form $(b_i, d_k)$ to distract policy iteration from switching the action $a_i$. Since every other state $b_j$ with $j \notin B$ has at least two actions $(b_j, d_k)$ with $k > 2i$, they move to the policy $\pi_{2i+2}^B$.

**Proposition 13.** *Policy iteration moves from the policy $\pi_{2i+1}^B$ to the policy $\pi_{R1}^B$.*

Since the action $a_i$ has been switched the value of the state $f_i$ is raised to $\text{Val}^{\pi_{R1}^B}(r_i) + (10n+4)(2^i - 2^{i-1}) - 4n$. The reward of $(10n+4)2^i$ is sufficiently large to cause policy iteration to switch the actions $(c_i, f_i)$ and $(x, f_i)$. It will also switch the actions $(b_j, f_i)$ where for every index $j < i$. Since every index $j \notin B$ other than $i$ has at least one action $(b_j, D_k)$, these states can be switched to the policy $\pi_{2i+3}^B(s)$. Therefore, we define:

$$\pi_{\text{R2}}^{B}(s) = \begin{cases} \pi_{0}^{B}(s) & \text{if } s = b_i \text{ or } s = r_i \text{ or } s \in \{c_j, b_j, r_j \ : \ j > i\}, \\ f_i & \text{if } s = x \text{ or } s = c_i \text{ or } s = b_j \text{ with } j < i, \\ \pi_{2i+3}^{B}(s) & \text{otherwise.} \end{cases}$$

The most important thing in this iteration is that every state $b_j$ with index $j < i$ is switched away from the action $a_j$. This provides the critical property of resetting every bit that has a smaller index than $i$. Another important property is that, while the action $(x, f_i)$ can be switched in this iteration, the action $(y, c_i)$ cannot be switched until after the action $(c_i, f_i)$ has been switched. This will provide a single iteration in which the value of $x$ will exceed the value of $y$, which is the first of the two conditions necessary to reset the deceleration lane.

**Proposition 14.** *Policy iteration moves from the policy $\pi_{R1}^{B}$ to the policy $\pi_{R2}^{B}$.*

In the next iteration the deceleration lane begins to reset as policy iteration switches $(d_k, x)$ for all $k$ and $(b_j, x)$ where $j > i$ and $j \notin B$. Policy iteration also switches $(y, c_i)$ and $(r_j, c_i)$ with $j < i$. We define:

$$\pi_{\text{R3}}^{B}(s) = \begin{cases} \pi_{0}^{B}(s) & \text{if } s \in \{c_j, b_j, r_j \ : \ j \geq i\} \cup \{x\}, \\ c_i & s = y \text{ or } s = r_j \text{ with } j < i, \\ x & s = d_k \text{ for some } k \text{ or } s = b_j \text{ with } j \notin B \setminus \{x\}, \\ \pi_{2i+3}^{B}(s) & \text{if } s = c_j \text{ with } j < i. \end{cases}$$

The switching of $(y, c_i)$ provides the second condition for the reset of the deceleration lane. After the action is switched the value of $y$ will be $\text{Val}^{\pi_{\text{R2}}^{B}}(f_i) + 4n + 1$ whereas the value of $x$ will be $\text{Val}^{\pi_{\text{R2}}^{B}}(f_i)$. Therefore, policy iteration will reset the deceleration lane in the next iteration. It is also important that the action $(b_j, x)$ for $j \notin B$ is switchable in this iteration, since if $i + 1 \notin B$ then $b_{i+1}$ will have run out of actions $(b_{i+1}, d_k)$ to distract it from switching $a_{i+1}$. The is the reason why each state $b_i$ must have $2i$ actions to the deceleration lane.

**Proposition 15.** *Policy iteration moves from the policy $\pi_{R2}^{B}$ to the policy $\pi_{R3}^{B}$.*

Finally, once policy iteration has reached the policy $\pi_{R3}^{B}$ it will move to the policy $\pi_{0}^{B'}$. This involves completing the reset of the deceleration lane by switching $(d_k, y)$ for all $k$, and switching the actions $(b_j, y)$ for every state $b_j$ with index $j \notin B'$. It also makes the final step in transforming the policy $\pi^{B}$ to the policy $\pi^{B'}$ by switching the actions $(c_j, r_j)$ at every state $c_j$ with $j < i$.

**Proposition 16.** *Policy iteration moves from the policy $\pi_{R3}^{B}$ to the policy $\pi_{0}^{B'}$.*

When combined with Proposition 12, the propositions in this section imply that policy iteration will move from the policy $\pi_{0}^{B}$ to the policy $\pi_{0}^{B'}$. The optimal policy for the example is $\pi_{2n+1}^{B}$ where $B = \{1, 2, \ldots n\}$. This is the policy that selects $a_i$ at $b_i$ for all $i$, and in which the deceleration lane has reached its optimal policy. Our results so far indicate that if we begin policy iteration at the policy $\pi_0^\emptyset$, then policy iteration must pass through a policy $\pi_0^B$ for every $B \subseteq \{1, 2, \ldots n\}$. Therefore, it will take at least $2^n$ iterations to terminate.

**Theorem 17.** *When policy iteration for the total reward criterion is applied to the policy $\pi_0^\emptyset$ it will take at least $2^n$ iterations to find the optimal policy.*

Finally, we can also argue that the example also provides an exponential lower bound for policy iteration for the average-reward criterion. The first thing to note that $G^\pi(s) = 0$ for every policy that we have specified. This is because all runs eventually reach the sink state $c_{n+1}$. Since the reward of the action $(c_{n+1}, c_{n+1})$ is 0, the long term average-reward of every policy will be 0. Note that when $G^\pi(S) = 0$, the bias optimality equation (3) becomes identical to the total reward optimality equation (1). This causes policy iteration for the average-reward criterion to behave identically to policy iteration for the total reward criterion on this example.

**Theorem 18.** *When policy iteration for the average-reward criterion is applied to the policy $\pi_0^\emptyset$ it will take at least $2^n$ iterations to find the optimal policy.*

**Acknowledgements.** I am indebted to Marcin Jurdziński for his guidance, support, and encouragement while preparing this paper.

## A    Proofs for Section 4.1

*Proof (of Proposition 2).* Using the optimality equation given in (4) we get:

$$\text{Val}^\pi(b_i) = \text{r}(b_i, a_i) + \sum_{s \in S} p(s|b_i, a_i) \text{Val}^\pi(s)$$

$$\text{Val}^\pi(b_i) = 0 + (1 - \frac{1}{(10n+4)2^n}) \text{Val}^\pi(b_i) + \frac{\text{Val}^\pi(g_i)}{(10n+4)2^n}$$

$$\frac{\text{Val}^\pi(b_i)}{(10n+4)2^n} = \frac{\text{Val}^\pi(g_i)}{(10n+4)2^n}$$

$$\text{Val}^\pi(b_i) = \text{Val}^\pi(g_i)$$

□

*Proof (of Proposition 4).* We have:

$$\text{Appeal}(b_i, a_i) = \text{r}(b_i, a_i) + \sum_{s \in S} p(s|b_i, a_i) \text{Val}^\pi(s)$$

$$= 0 + (1 - \frac{2^{-(n)}}{10n+4}) \text{Val}^\pi(b_i) + \frac{2^{-(n)}}{10n+4} \text{Val}^\pi(g_i)$$

$$< \text{Val}^\pi(b_i) + \frac{2^{-(n)}}{10n+4} \text{Val}^\pi(g_i)$$

$$\leq \text{Val}^\pi(b_i) + \frac{2^{-(n)}}{10n+4} \cdot (10n+4)2^n$$

$$= \text{Val}^\pi(b_i) + 1$$

□

## B    Proof of Proposition 6

**Proposition 19.** *When $j \leq i + 1$ we have*

$$\text{Appeal}^{\pi_i}(d_j, d_{j-1}) = \text{Val}^{\pi_i}(y) + 4n - j + 1.$$

*Proof.* By the definition of $\pi_i$ we have that $\pi_i(d_j) = d_{j-1}$ for all states $d_j$ with $j \leq i$, and we have that $\pi_i(z) = y$. Using the definition of appeal, and applying the optimality equation (4) repeatedly gives, for every action $(d_j, d_{j-1})$ with $j \leq i+1$:

$$\text{Appeal}^{\pi_i}(d_j, d_{j-1}) = \text{r}(d_j, d_{j-1}) + \text{Val}^{\pi_i}(d_j)$$

$$= \sum_{k=1}^{j} \text{r}(d_k, d_{k-1}) + \text{r}(z, y) + \text{Val}^{\pi_i}(y)$$

$$= -j + (4n+1) + \text{Val}^{\pi_i}(y)$$

□

**Proposition 20.** *When $i+1 < j \leq 2n$ we have $\mathrm{Appeal}^{\pi_i}(d_j, d_{j-1}) = \mathrm{Val}^{\pi_i}(y) - 1$.*

*Proof.* By definition we have that $\pi_i(d_{j-1}) = y$. Using the definition of appeal and the optimality equation gives:

$$\begin{aligned}\mathrm{Appeal}^{\pi_i}(d_j, d_{j-1}) &= \mathrm{r}(d_j, d_{j-1}) + \mathrm{Val}^{\pi_i}(d_{j-1}) \\ &= \mathrm{r}(d_j, d_{j-1}) + \mathrm{r}(d_{j-1}, d_y) + \mathrm{Val}^{\pi_i}(y) \\ &= \mathrm{Val}^{\pi_i}(y) - 1\end{aligned}$$

□

**Proposition 21.** *If $\mathrm{Val}^{\pi_i}(x) < \mathrm{Val}^{\pi_i}(y)$ then $\mathrm{Appeal}^{\pi_i}(d_j, x) < \mathrm{Appeal}^{\pi_i}(d_j, y)$ for all $j$. If $\mathrm{Val}^{\pi_i}(x) > \mathrm{Val}^{\pi_i}(y)$ then $\mathrm{Appeal}^{\pi_i}(d_j, x) > \mathrm{Appeal}^{\pi_i}(d_j, y)$ for all $j$.*

*Proof.* Using the definition of appeal for the state $d_j$ gives two equalities:

$$\begin{aligned}\mathrm{Appeal}^{\pi_i}(d_j, y) &= \mathrm{r}(d_j, y) + \mathrm{Val}^{\pi_i}(y) \\ \mathrm{Appeal}^{\pi_i}(d_j, x) &= \mathrm{r}(d_j, x) + \mathrm{Val}^{\pi_i}(x)\end{aligned}$$

Observe that for all $j$ we have $\mathrm{r}(d_j, y) = \mathrm{r}(d_j, x)$. Therefore we can conclude that when $\mathrm{Val}^{\pi_i}(x) < \mathrm{Val}^{\pi_i}(y)$ we have $\mathrm{Appeal}^{\pi_i}(d_j, x) < \mathrm{Appeal}^{\pi_i}(d_j, y)$, and when $\mathrm{Val}^{\pi_i}(x) > \mathrm{Val}^{\pi_i}(y)$ we have $\mathrm{Appeal}^{\pi_i}(d_j, x) > \mathrm{Appeal}^{\pi_i}(d_j, y)$. □

*Proof (of Proposition 6).* To prove this claim it is sufficient to show that policy iteration moves from the policy $\pi_j$ to the policy $\pi_{j+1}$. We break this task into two parts: we will first show that $(d_k, d_{k-1})$ is the most appealing action at every state $d_k$ with $1 \leq k \leq i+1$, and then we will show that $(d_k, y)$ is the most appealing action at every other state. In both cases, however, Assumption 5 combined with Proposition 21 implies that $\mathrm{Appeal}^{\pi_j}(d_k, x) < \mathrm{Appeal}^{\pi_j}(d_k, y)$. Therefore, the action $(d_k, x)$ can be ignored.

For a state $d_k$ with $1 \leq k \leq i+1$, Proposition 19 combined with the fact that $j \leq 2n$ gives:

$$\begin{aligned}\mathrm{Appeal}^{\pi_j}(d_k, d_{k-1}) &= \mathrm{Val}^{\pi_j}(y) + 4n - k + 1 \\ &\geq \mathrm{Val}^{\pi_j}(y) + 2n + 1 \\ &> \mathrm{Val}^{\pi_j}(y) = \mathrm{Appeal}^{\pi_j}(d_k, y)\end{aligned}$$

Therefore $(d_k, d_{k-1})$ is the most appealing action at these states.

We now consider the other states. The state $d_0$ only has the actions $(d_0, y)$ and $(d_0, x)$, and so this state can be ignored. The remaining states are those states $d_k$ with $k$ in the range $i + 1 < k \leq 2i$, and these states have an additional action $(d_k, d_{k-1})$. Proposition 19 gives:

$$\mathrm{Appeal}^{\pi_j}(d_k, d_{k-1}) = \mathrm{Val}^{\pi_j}(y) - 1 < \mathrm{Val}^{\pi_j}(y) = \mathrm{Appeal}^{\pi_j}(d_k, y).$$

Therefore $(d_k, y)$ is the most appealing action at these states. □

## C  Proof of Proposition 8

**Proposition 22.** *If $\pi$ is either $\pi_j^o$ or $\pi_j^c$, then when $1 \leq k \leq j$ we have $\mathrm{Appeal}^\pi(b_i, d_k) = \mathrm{Val}^\pi(y) + 4n + k + 1$.*

*Proof.* Since $\pi(d_k) = \pi_j(d_k)$ for every state $d_k$, we can apply Proposition 19 to give:
$$\begin{aligned}\mathrm{Appeal}^\pi(b_i, d_k) &= \mathrm{r}(b_i, d_k) + \mathrm{Val}^\pi(d_k)\\ &= 2k + (\mathrm{Val}^\pi(y) + 4n - k + 1)\\ &= \mathrm{Val}^\pi(y) + 4n + k + 1\end{aligned}$$
□

**Proposition 23.** *If $\pi$ is either $\pi_j^o$ or $\pi_j^c$, then when $j < k \leq 2i$ we have $\mathrm{Appeal}^\pi(b_i, d_k) = \mathrm{Val}^\pi(y) + 2k$.*

*Proof.* Using the definition of appeal, and the fact that $\pi(d_k) = y$ when $1 < k \leq 2n$ gives:
$$\begin{aligned}\mathrm{Appeal}^\pi(b_i, d_k) &= \mathrm{r}(b_i, d_k) + \mathrm{Val}^\pi(d_k)\\ &= \mathrm{r}(b_i, d_k) + \mathrm{r}(d_k, y) + \mathrm{Val}^\pi(y)\\ &= 2k + \mathrm{Val}^\pi(y)\end{aligned}$$
□

**Proposition 24.** *Suppose that Assumptions 3 and 5 hold. When policy iteration is applied to $\pi_0^o$ it will move to the policy $\pi_1^o$.*

*Proof.* To prove this proposition we must show that $(b_i, d_{2i})$ is the most appealing action at the state $b_i$ in the policy $\pi_0^o$. For the action $(b_i, x)$, Assumption 5 implies:
$$\mathrm{Appeal}^{\pi_0^o}(b_i, x) = \mathrm{Val}^{\pi_0^o}(x) < \mathrm{Val}^{\pi_0^o}(y) = \mathrm{Appeal}^{\pi_0^o}(b_i, y)$$

For the action $(b_i, y)$, Proposition 23 combined with the fact that $4i > 1$ gives:
$$\mathrm{Appeal}^{\pi_0^o}(b_i, y) = \mathrm{Val}^{\pi_0^o}(y) + 1 < \mathrm{Val}^{\pi_0^o}(y) + 4i = \mathrm{Appeal}^{\pi_0^o}(b_i, d_{2i})$$

Therefore, the action $(b_i, d_{2i})$ is more appealing than the actions $(b_i, x)$ and $(b_i, y)$. For the actions $(b_i, d_k)$ with $k$ in the range $1 \leq k < 2i$, Proposition 23, and the fact that $k < 2i$ give:
$$\mathrm{Appeal}^{\pi_0^o}(b_i, d_k) = \mathrm{Val}^{\pi_0^o}(y) + 2k < \mathrm{Val}^{\pi_0^o}(y) + 4i = \mathrm{Appeal}^{\pi_0^o}(b_i, d_{2i})$$

Finally, for the action $(b_i, a_i)$, Proposition 4 combined with the fact that $i \geq 1$ give:
$$\begin{aligned}\mathrm{Appeal}^{\pi_0^o}(b_i, a_i) &< \mathrm{Val}^{\pi_0^o}(b_i) + 1\\ &= \mathrm{Val}^{\pi_0^o}(y) + 1\\ &< \mathrm{Val}^{\pi_0^o}(y) + 4i = \mathrm{Appeal}^{\pi_0^o}(b_i, d_{2i})\end{aligned}$$

Therefore $(b_i, d_{2i})$ is the most appealing action at $b_i$ in the policy $\pi_0^o$. □

*Proof (of Proposition 8).* The fact that policy iteration moves from $\pi_0^o$ to $\pi_1^o$ was shown in Proposition 24. To complete the proof we must show that policy iteration moves from $\pi_j^o$ to $\pi_{j+1}^o$ for $j$ in the range $1 \le j < 2i$. Proposition 6 implies that this holds for the states $d_k$, which means that we are only concerned with the action chosen at the state $b_i$. For this state we must show that $(b_i, d_j)$ is the most appealing action at the state $b_i$ in the policy $\pi_j$. Assumption 7 implies that the actions $(b_i, f_j)$ can not be switched by policy iteration.

For the action $(b_i, x)$, Assumption 5 implies:

$$\text{Appeal}^{\pi_j^o}(b_i, x) = \text{Val}^{\pi_j^o}(x) < \text{Val}^{\pi_j^o}(y) = \text{Appeal}^{\pi_j^o}(b_i, y).$$

For the action $(b_i, y)$ the fact that $j > 0$ and Proposition 22 give:

$$\text{Appeal}^{\pi_j^o}(b_i, y) = \text{Val}^{\pi_j^o}(y) + 1 < \text{Val}^{\pi_j^o}(y) + 4n + j + 1 = \text{Appeal}^{\pi_j^o}(b_i, d_j)$$

Therefore the action $(b_i, d_j)$ is more appealing than the actions $(b_i, x)$ and $(b_i, y)$.

For the actions of the form $(b_i, d_k)$ we consider two cases. For states $d_k$ with $1 \le k < j$ we have by Proposition 22, and the fact that $k < j$:

$$\text{Appeal}^{\pi_j^o}(b_i, d_k) = \text{Val}^{\pi_j^o}(y) + 4n + k + 1$$
$$< \text{Val}^{\pi_j^o}(y) + 4n + j + 1 = \text{Appeal}^{\pi_j^o}(b_i, d_j)$$

For states $d_k$ with $j < k < 2i$ we have by Proposition 23, and the fact that $2k \le 4n$:

$$\text{Appeal}^{\pi_j^o}(b_i, d_k) = \text{Val}^{\pi_j^o}(y) + 2k$$
$$< \text{Val}^{\pi_j^o}(y) + 4n + j + 1 = \text{Appeal}^{\pi_j^o}(b_i, d_j)$$

Therefore the action $(b_i, d_j)$ is more appealing than the actions $(b_i, d_k)$ for every $k \ne j$.

Finally, we consider the action $(b_i, a_i)$. Using Proposition 4 and Proposition 22 gives:

$$\text{Appeal}^{\pi_j^o}(b_i, a_i) < \text{Val}^{\pi_j^o}(b_i) + 1 = \text{Appeal}^{\pi_j^o}(b_i, d_{j-1}) + 1$$
$$= \text{Val}^{\pi_j^o}(y) + 4n + j + 1 = \text{Appeal}^{\pi_j^o}(b_i, d_j)$$

Therefore the action $(b_i, d_j)$ is more appealing than the action $(b_i, a_i)$. We have now shown that the action $(b_i, d_j)$ is more appealing than every other action at $b_i$ in the policy $\pi_j^o$. □

## D  Proof of Proposition 10

*Proof (of Proposition 10).* For the states $d_k$ this proposition follows from Proposition 6. To complete the proof we must show that $(b_i, a_i)$ is the most appealing action at $b_i$ in the policy $\pi_j^c$. Assumption 7 implies that the actions $(b_i, f_j)$ can not be switched by policy iteration.

For the action $(b_i, x)$, Assumption 5 implies:

$$\mathrm{Appeal}^{\pi_j^c}(b_i, x) = \mathrm{Val}^{\pi_j^c}(x) < \mathrm{Val}^{\pi_j^c}(y) = \mathrm{Appeal}^{\pi_j^c}(b_i, y)$$

Furthermore, for the action $(b_i, y)$, Assumption 9 implies:

$$\mathrm{Appeal}^{\pi_j^c}(b_i, y) = \mathrm{Val}^{\pi_j^c}(y) + 1 < \mathrm{Val}^{\pi_j^c}(b_i) = \mathrm{Appeal}^{\pi_j^c}(b_i, a_i)$$

Therefore, the action $(b_i, a_i)$ is more appealing than the actions $(b_i, x)$ and $(b_i, y)$.

For the actions $(b_i, d_k)$ we consider two cases. Firstly, when $1 \leq k \leq j$ we can apply Proposition 22, the fact that $k \leq 2n$, and Assumption 9 to give:

$$\mathrm{Appeal}^{\pi_j^c}(b_i, d_k) \leq \mathrm{Val}^{\pi_j^c}(y) + 4n + k + 1$$
$$\leq \mathrm{Val}^{\pi_j^c}(y) + 6n + 1 < \mathrm{Val}^{\pi_j^c}(b_i) = \mathrm{Appeal}^{\pi_j^c}(b_i, a_i)$$

When $j < k \leq 2i$, Proposition 23, the fact that $k \leq 2n$, and Assumption 9 give:

$$\mathrm{Appeal}^{\pi_j^c}(b_i, d_k) \leq \mathrm{Val}^{\pi_j^c}(y) + 2k$$
$$\leq \mathrm{Val}^{\pi_j^c}(y) + 4n < \mathrm{Val}^{\pi_j^c}(b_i) = \mathrm{Appeal}^{\pi_j^c}(b_i, a_i)$$

Therefore, the action $(b_i, a_i)$ is the most appealing action at the state $b_i$ in the policy $\pi_j^c$. □

## E  Proof of Proposition 11

**Proposition 25.** *Let $B$ be a configuration and $\pi$ be a member of* $\mathrm{Sequence}(B)$. *For every $i$ we have:*

$$\mathrm{Val}^\pi(c_i) = \begin{cases} \sum_{j \in B^{\geq i}} (10n+4)(2^j - 2^{j-1}) & \text{if } i \in B, \\ \sum_{j \in B^{\geq i}} (10n+4)(2^j - 2^{j-1}) - 1 & \text{otherwise.} \end{cases}$$

*Proof.* We first consider the case where $i \in B$. If $k = \min(B \cup \{n+1\})$ then the definition of $\pi$, and Proposition 2 give:

$$\mathrm{Val}^\pi(c_i) = \mathrm{r}(c_i, f_i) + \mathrm{r}(f_i, b_i) + \mathrm{Val}^\pi(b_i)$$
$$= \mathrm{r}(c_i, f_i) + \mathrm{r}(f_i, b_i) + \mathrm{r}(g_i, r_i) + \mathrm{r}(r_i, c_k) + \mathrm{Val}^\pi(c_k)$$
$$= (4n+1) - ((10n+4)2^{i-1} - 4n) + (10n+4)2^i - 1 + \mathrm{Val}^\pi(c_k)$$
$$= (10n+4)(2^i - 2^{i-1}) + \mathrm{Val}^\pi(c_k)$$

If $k = n+1$ then we are done because $\mathrm{Val}^\pi(c_{n+1}) = 0$. Otherwise, repeated substitution of the above expression for $\mathrm{Val}^\pi(c_k)$ gives:

$$\mathrm{Val}^\pi(c_i) = \sum_{j \in B^{\geq i}} (10n+4)(2^j - 2^{j-1}) + \mathrm{Val}^\pi(c_{j+1})$$
$$= \sum_{j \in B^{\geq i}} (10n+4)(2^j - 2^{j-1})$$

We now consider the case where $i \notin B$. The definition of $\pi$ gives:

$$\begin{aligned}\operatorname{Val}^\pi(c_i) &= \operatorname{r}(c_i, r_i) + \operatorname{r}(r_i, c_j) + \operatorname{Val}^\pi(c_j) \\ &= \operatorname{Val}^\pi(c_j) - 1 \\ &= \sum_{j \in B^{\geq i}} (10n + 4)(2^j - 2^{j-1}) - 1\end{aligned}$$

$\square$

**Proposition 26.** *Let $B$ be a configuration and $\pi$ be a member of* $\operatorname{Sequence}(B)$. *For every $i \in B$ and $j \in B$ such that $j > i$, we have $\operatorname{Val}^\pi(c_i) < \operatorname{Val}^\pi(c_j) + (10n + 4)(2^{j-1} - 2^{i-1})$.*

*Proof.* Let $C = B^{\geq i} \cap B^{<j}$ be the members of $B$ that lie between indices $i$ and $j - 1$. Using Proposition 25 gives:

$$\begin{aligned}\operatorname{Val}^\pi(c_i) &= \sum_{k \in B^{\geq i}} (10n + 4)(2^k - 2^{k-1}) \\ &= \sum_{k \in C} (10n + 4)(2^k - 2^{k-1}) + \operatorname{Val}^\pi(c_j)\end{aligned}$$

We use the fact that $(10n + 4)(2^k - 2^{k-1}) > 0$ for all $k$ and the fact that $i > 0$ to obtain:

$$\begin{aligned}\sum_{k \in C} (10n + 4)(2^k - 2^{k-1}) &\leq \sum_{k=i}^{j-1} (10n + 4)(2^k - 2^{k-1}) \\ &= (10n + 4)(2^{j-1} - 2^{i-1})\end{aligned}$$

Therefore, we have:

$$\operatorname{Val}^\pi(c_i) < \operatorname{Val}^\pi(c_j) + (10n + 4)(2^{j-1} - 2^{i-1})$$

$\square$

**Proposition 27.** *Let $B$ be a configuration and $\pi$ be a member of* $\operatorname{Sequence}(B)$. *For every $i \in B$ and $j \in B$ such that $j \geq i$, we have $\operatorname{Val}^\pi(c_i) \geq \operatorname{Val}^\pi(c_j)$.*

*Proof.* Let $C = B^{\geq i} \cap B^{<j}$ be the members of $B$ that lie between indices $i$ and $j - 1$. Proposition 25 and the fact that $2^j - 2^{j-1}$ is positive for every $j$ imply:

$$\operatorname{Val}^\pi(c_i) = \sum_{j \in C} (10n + 4)(2^j - 2^{j-1}) + \operatorname{Val}^\pi(c_j) \geq \operatorname{Val}^\pi(c_j)$$

$\square$

**Proposition 28 (Proof of Assumption 3).** *For every configuration $B$ and every policy $\pi$ in* $\operatorname{Sequence}(B)$ *we have $\operatorname{Val}^\pi(b_i) > 0$ and $\operatorname{Val}^\pi(g_i) \leq (10n+4)2^n$, for all $i$.*

*Proof.* To prove that $\text{Val}^\pi(b_i) > 0$ we must consider two cases. When $i \in B$ and $k = \min(B \cup \{n+1\})$ then we can apply Proposition 2 to obtain:

$$\text{Val}^\pi(b_i) = \text{Val}^\pi(g_i) = \text{Val}^\pi(r_i) + (10n+4)2^i$$
$$= \text{Val}^\pi(c_k) + (10n+4)2^i - 1$$

Since $i > 0$ we have $(10n+4)2^i - 1 > 0$, and we must therefore argue that $\text{Val}^\pi(c_k) \geq 0$. If $k = n+1$ then we are done because $\text{Val}^\pi(c_{n+1}) = 0$. Otherwise, we can apply Proposition 25 to give:

$$\text{Val}^\pi(c_k) = \sum_{j \in B^{\geq i}} (10n+4)(2^j - 2^{j-1})$$

This summation is clearly positive, since $(10n+4)(2^j - 2^{j-1}) > 0$ for every $j$.

We will now show that $\text{Val}^\pi(b_i) > 0$ in the case where $i \notin B$. In this case we have $\pi(b_i) = d_k$ for some $k$, and that $\pi(d_l) = d_{l-1}$ for all $l$ in the range $1 \leq l \leq k$. We can therefore apply Proposition 19 and Proposition 26 to give:

$$\text{Val}^\pi(b_i) = 4n + k + 1 + \text{Val}^\pi(y)$$
$$= 4n + k + 1 + \sum_{j \in B^{\geq i}} (10n+4)(2^j - 2^{j-1})$$

Since $k > 0$ we have that $(4n + k + 1) > 0$, and we have already argued that the summation will be non-negative. This implies that the entire expression will be positive. Therefore, we have shown that $\text{Val}^\pi(b_i) > 0$ for all $i$.

Finally, we argue that $\text{Val}^\pi(g_i) \leq (10n+4)2^n$. If $k = \min(B \cup \{n+1\})$ then we have:

$$\text{Val}^\pi(g_i) = (10n+4)2^i + \text{Val}^\pi(r_i) = (10n+4)2^i - 1 + \text{Val}^\pi(c_k)$$

If $k = n+1$ then we are done because $\text{Val}^\pi(c_{n+1}) = 0$ and $(10n+4)2^i - 1 < (10n+4)2^n$ for all $i \leq n$. Otherwise, we can apply Proposition 25 and the fact that $k - 1 \geq i$ to obtain:

$$\text{Val}^\pi(g_i) \leq (10n+4)2^i - 1 + \text{Val}^\pi(c_{n+1}) + (10n+4)(2^n - 2^{k-1})$$
$$\leq \text{Val}^\pi(c_{n+1}) + (10n+4)2^n - 1$$
$$\leq (10n+4)2^n$$

□

**Proposition 29 (Proof of Assumption 5).** *For every configuration $B$ and every policy $\pi$ in $\text{Sequence}(B)$ we have $\text{Val}^\pi(y) > \text{Val}^\pi(x)$.*

*Proof.* We first consider the case where $B = \emptyset$. In this situation the definition of $\pi$ implies that $\pi(y) = c_{n+1}$ and $\pi(x) = c_{n+1}$. Therefore, we have:

$$\text{Val}^\pi(x) = -1 + \text{Val}^\pi(c_{n+1}) < \text{Val}^\pi(c_{n+1}) = \text{Val}^\pi(y)$$

We now consider the case where $B \neq \emptyset$. By definition of $\pi$ we have that there is some index $i \in B$ such that $\pi(y) = c_i$ and $\pi(x) = f_i$. Moreover, since $i \in B$ we have that $\pi(c_i) = f_i$. We therefore have two equalities:

$$\text{Val}^\pi(y) = \text{r}(y, c_i) + \text{r}(c_i, f_i) + \text{Val}^\pi(f_i) = \text{Val}^\pi(f_i) + 4n + 1$$
$$\text{Val}^\pi(x) = \text{r}(x, f_i) + \text{Val}^\pi(f_i) = \text{Val}^\pi(f_i)$$

Clearly, since $4n + 1 > 0$ we have $\text{Val}^\pi(y) > \text{Val}^\pi(x)$. $\square$

**Proposition 30 (Proof of Assumption 7).** *Let $B$ be a configuration and $\pi$ be a member of $\text{Sequence}(B)$. If $(b_i, f_j)$ is an action such that $j \in B$ then $\text{Appeal}^\pi(b_i, f_j) < \text{Val}^\pi(b_i)$.*

*Proof.* To prove this proposition we must consider four cases. Firstly, when $j \in B$ and $i \notin B$ we can apply Proposition 22, the fact that $k \geq 0$, the fact that $\min(B) \leq j$, and Proposition 27 to give:

$$\text{Val}^\pi(b_i) = \text{Val}^\pi(y) + 4n + k + 1 \geq \text{Val}^\pi(c_{\min(B)}) + 4n + 1$$
$$\geq \text{Val}^\pi(c_j) + 4n + 1$$
$$= \text{Val}^\pi(f_j) + 8n + 2$$
$$> \text{Val}^\pi(f_j) + 4n + 1 = \text{Appeal}^\pi(b_i, f_j)$$

Secondly, we consider the case where $j \in B$ and $i \in B$. In this case we can apply Proposition 2, the fact that $\min(B^{\geq i}) \leq j$, Proposition 27, and the fact that $i > 0$ to obtain:

$$\text{Val}^\pi(b_i) = (10n + 4)2^i + \text{Val}^\pi(r_i)$$
$$= (10n + 4)2^i - 1 + \text{Val}^\pi(c_{\min(B^{\geq i})})$$
$$\geq (10n + 4)2^i - 1 + \text{Val}^\pi(c_j)$$
$$= (10n + 4)2^i + 4n + \text{Val}^\pi(f_j)$$
$$> \text{Val}^\pi(f_j) + 4n + 1 = \text{Appeal}^\pi(b_i, f_j)$$

Thirdly, we consider the case where $j \notin B$ and $i \notin B$. In this case, the fact that $\pi(b_i) = \pi(b_j)$ gives:

$$\text{Appeal}^\pi(b_i, f_j) = 4n + 1 + \text{Val}^\pi(f_j)$$
$$= -(10n + 4)2^{j-1} + 1 + \text{Val}^\pi(b_j)$$
$$= -(10n + 4)2^{j-1} + 1 + \text{Val}^\pi(\pi(b_j))$$
$$= -(10n + 4)2^{j-1} + 1 + \text{Val}^\pi(\pi(b_i)) < \text{Val}^\pi(\pi(b_i))$$

Finally, we consider the case where $j \notin B$ and $i \in B$. Proposition 2, and the fact that $k \leq 2n$, imply:

$$\text{Appeal}^\pi(b_i, f_j) = -(10n + 4)2^j + 1 + \text{Val}^\pi(b_j)$$
$$= -(10n + 4)2^j + 4n + k + 2 + \text{Val}^\pi(y)$$
$$\leq -(10n + 4)2^j + 6n + 2 + \text{Val}^\pi(c_{\min(B)})$$

Let $l = \min(B^{>j} \cup \{n+1\})$ be the smallest bit in the configuration that is larger than $j$. By Proposition 25, and the fact that there is no bit in the configuration with an index $m$ in the range $j \leq m < l$, we have:

$$\mathrm{Val}^\pi(c_{\min(B)}) = \sum_{j \in B^{<j}} (10n+4)(2^j - 2^{j-1}) + \mathrm{Val}^\pi(c_l)$$
$$\leq \mathrm{Val}^\pi(c_l) + (10n+4)(2^{j-1} - 2^0)$$

Therefore, we have:

$$\mathrm{Appeal}^\pi(b_i, f_j) \leq \mathrm{Val}^\pi(c_l) + (10n+4)(2^{j-1} - 1 - 2^j) + 6n + 2$$
$$\leq \mathrm{Val}^\pi(c_l) + (10n+4)(2^{j-1} - 2^j)$$
$$< \mathrm{Val}^\pi(c_l)$$

However, Proposition 2, Proposition 27, and the fact that $i \geq 0$ imply:

$$\mathrm{Val}^\pi(b_i) = (10n+4)2^i + \mathrm{Val}^\pi(r_i) = (10n+4)2^i - 1 + \mathrm{Val}^\pi(c_{\min(B^{\geq i})})$$
$$\geq (10n+4)2^i - 1 + \mathrm{Val}^\pi(c_l)$$
$$\geq \mathrm{Val}^\pi(c_l)$$

$\square$

**Proposition 31 (Proof of Assumption 9).** *For every configuration $B$ and every policy $\pi$ in $\mathrm{Sequence}(B)$ we have $\mathrm{Val}^\pi(y) + 6n + 1 < \mathrm{Val}^\pi(b_i)$, for every index $i \in B$.*

*Proof.* If $B = \emptyset$ then there are no indices $i \in B$, and so the proposition is vacuously true. Otherwise, the definition of $\pi$ implies that $\pi(y) = c_{\min(B)}$. Applying Proposition 25 gives:

$$\mathrm{Val}^\pi(y) = \mathrm{Val}^\pi(c_{\min(B)})$$
$$\leq \mathrm{Val}^\pi(c_i) + (10n+4)(2^{i-1} - 2^{\min(B)-1})$$

Since $i \in B$ we have $\pi(c_i) = f_i$. Therefore we can apply the optimality equation, Proposition 26, and the fact that $\min(B) - 1 \geq 0$ to obtain:

$$\mathrm{Val}^\pi(y) \leq \mathrm{Val}^\pi(c_i) + (10n+4)(2^{i-1} - 2^{\min(B)-1})$$
$$\leq \mathrm{Val}^\pi(c_i) + (10n+4)(2^{i-1} - 2^0)$$
$$= \mathrm{Val}^\pi(f_i) + (4n+1) + (10n+4)(2^{i-1} - 1)$$
$$= \mathrm{Val}^\pi(b_i) - (10n+4)2^{i-1} - 4n + (4n+1) + (10n+4)(2^{i-1} - 1)$$
$$= \mathrm{Val}^\pi(b_i) - (10n+3)$$

It is now clear that $\mathrm{Val}^\pi(y) + 6n + 1 < \mathrm{Val}^\pi(b_i)$.

$\square$

# F  Proof of Proposition 12

**Proposition 32.** *Let $B$ be a configuration and $\pi$ be a member of $\mathrm{Sequence}(B)$. If there are two indices $i \in B \cup \{n+1\}$ and $j \in B \cup \{n+1\}$ such that $i < j$, then $\mathrm{Val}^\pi(c_i) > \mathrm{Val}^\pi(c_j)$.*

*Proof.* We first prove the statement for the states $c_i$ and $c_j$. By Proposition 25, we have:
$$\mathrm{Val}^\pi(c_k) = \sum_{l \in B^{\geq k}} (10n+4)(2^l - 2^{l-1}).$$

Since $(10n+4)(2^i - 2^{i-1}) > 0$ we have:
$$\begin{aligned}
\mathrm{Val}^\pi(c_j) &= \sum_{l \in B^{\geq j}} (10n+4)(2^l - 2^{l-1}) \\
&< \sum_{l \in B^{\geq j}} (10n+4)(2^l - 2^{l-1}) + (10n+4)(2^i - 2^{i-1}) \\
&\leq \sum_{l \in B^{\geq i}} (10n+4)(2^l - 2^{l-1}) = \mathrm{Val}^\pi(c_i)
\end{aligned}$$

□

**Proposition 33.** *Let $B$ be a configuration and $\pi$ be a member of $\mathrm{Sequence}(B)$. For every subset $C \subseteq B \cup \{n+1\}$, if $i = \min(C)$ then we have $\mathrm{Val}^\pi(c_i) > \mathrm{Val}^\pi(c_j)$ for every $j \neq i$.*

*Proof.* Proposition 32 implies that the state $c_i$ has a higher value than every other state $c_j$ with $j \in C$. To complete the proof we must eliminate the states $c_j$ with $j \notin C$. We will accomplish this by arguing that for every such state $c_j$ there is some index $k \in C$ such that $\mathrm{Val}^\pi(c_k) > \mathrm{Val}^\pi(c_j)$. We choose $k = \min(C^{>j} \cup \{n+1\})$ to be the smallest index in $C$ that is higher than $j$, or the index of the sink if $j$ is the largest index in $C$. Since $j \notin C$ we have:
$$\begin{aligned}
\mathrm{Val}^\pi(c_j) &= \mathrm{r}(c_j, r_j) + \mathrm{r}(r_j, c_k) + \mathrm{Val}^\pi(c_k) \\
&= \mathrm{Val}^\pi(c_k) - 1 < \mathrm{Val}^\pi(c_k)
\end{aligned}$$

□

**Proposition 34.** *Let $B$ be a configuration and $\pi$ be a member of $\mathrm{Sequence}(B)$. If $i \notin B$ we have $\mathrm{Val}^\pi(f_i) + 4n + 1 < \mathrm{Val}^\pi(c_k)$ where $k = \min(B^{>i} \cup n+1)$.*

*Proof.* Using Proposition 22, and the fact that $i \leq n$, to obtain the value of $f_i$ gives:
$$\begin{aligned}
\mathrm{Val}^\pi(f_i) &= -(10n+4)2^{i-1} - 4n + 6n + 1 + \mathrm{Val}^\pi(y) \\
&= -(10n+4)2^{i-1} + 2n + 1 + \mathrm{Val}^\pi(y)
\end{aligned}$$

Using Proposition 25 to obtain the value of $y$ in terms of the state $c_k$ gives:

$$\text{Val}^\pi(f_i) = -(10n+4)2^{i-1} + 2n + 1 + \sum_{j \in B^{<i}} (10n+4)(2^j - 2^{j-1}) + \text{Val}^\pi(c_k)$$

$$\leq -(10n+4)2^{i-1} + 2n + 1 + \sum_{j=0}^{i-1}(10n+4)(2^j - 2^{j-1}) + \text{Val}^\pi(c_k)$$

$$= -(10n+4)2^{i-1} + 2n + 1 + (10n+4)(2^{i-1} - 2^0) + \text{Val}^\pi(c_k)$$

$$= 2n + 1 + -(10n+4) + \text{Val}^\pi(c_k)$$

$$= -8n - 3 + \text{Val}^\pi(c_k)$$

Therefore $\text{Val}^\pi(f_i) + 4n + 1 < \text{Val}^\pi(c_k)$. □

**Proposition 35.** *Let $B$ be a configuration and $\pi$ be a member of* $\text{Sequence}(B)$. *If $i = \min(B \cup \{n+1\})$ then we have $\text{Val}^\pi(f_i) > \text{Val}^\pi(f_j)$ for every $j \neq i$.*

*Proof.* We begin by arguing that $\text{Val}^\pi(f_i) > \text{Val}^\pi(f_j)$ for the case where $j \in B$. Since $\text{Val}^\pi(c_k) = \text{Val}^\pi(f_k) + 4n + 1$ for every $k \in B$, we can apply Proposition 32 to obtain:

$$\text{Val}^\pi(f_i) = \text{Val}^\pi(c_i) - 4n - 1 > \text{Val}^\pi(c_j) - 4n - 1 = \text{Val}^\pi(f_j)$$

We now complete the proof by arguing that for every state $f_j$ with $j \notin B$ there is some state $f_k$ with $k \in B$ such that $\text{Val}^\pi(f_k) > \text{Val}^\pi(f_j)$. Proposition 34 implies that $\text{Val}^\pi(f_i) + 4n + 1 < \text{Val}^\pi(c_k)$ where $k = \min(B \cup n+1)$. Therefore:

$$\text{Val}^\pi(f_j) < \text{Val}^\pi(c_k) - (4n+1) = \text{Val}^\pi(f_k)$$

□

**Proposition 36.** *Let $B$ be a configuration and $\pi$ be a member of* $\text{Sequence}(B)$. *The state $c_i$ will not be switched away from $\pi(c_i)$.*

*Proof.* First we will consider the case where $i \in B$, where we must show that the state $c_i$ does not switch away from the action $(c_i, f_i)$. In this case, the fact $2^i - 2^{i-1} > 0$ implies:

$$\text{Appeal}^\pi(c_i, f_i) = (10n+4)(2^i - 2^{i-1}) + 1 + \text{Val}^\pi(r_i)$$
$$> \text{Val}^\pi(r_i) = \text{Appeal}^\pi(c_i, r_i)$$

Therefore, policy iteration will not switch away from the action $(c_i, f_i)$.

Now we consider the case where $i \notin B$. In this case Proposition 34 implies that if $k = \min(B^{>i} \cup n+1)$ then:

$$\text{Appeal}^\pi(c_i, f_i) = 4n + 1 + \text{Val}^\pi(f_i) < \text{Val}^\pi(c_k) \leq \text{Val}^\pi(c_k) - 1$$

We also have that $\text{Val}^\pi(c_i) = \text{Val}^\pi(c_k) - 1$. Therefore, policy iteration will not switch away from the current action at $c_k$. □

*Proof (of Proposition 12).* Our previous proofs have shown that this proposition is true for every state not contained in the set $\{c_i, f_i, g_i, r_i : 1 \leq i \leq n\} \cup \{x, y\}$, which this proof will deal with. Note that since the states $f_i$ and $g_i$ have only one successor they can be ignored. The rest of this proof is dedicated to showing that the states $c_i$, $r_i$, $x$, and $y$ do not switch away from their current action for every policy $\pi$ in Sequence($B$). For the state $c_i$ this proposition is a consequence of Proposition 36.

For the state $r_i$, we must show that the most appealing action is $(r_i, c_j)$ where $j = \min(B^{>j} \cup \{n+1\})$. When $C = B^{>j} \cup \{n+1\}$, Proposition 33 implies that $\mathrm{Val}^\pi(c_j) > \mathrm{Val}^\pi(c_k)$ for every $j > k$. Since every outgoing action from $r_i$ has the same reward this also implies that the action $(r_i, c_k)$ is the most appealing action at $r_i$.

For the state $x$, we must show that the most appealing action is $(x, f_k)$, where $k = \min(B \cup n+1)$. Proposition 35 implies that $\mathrm{Val}^\pi(c_k) > \mathrm{Val}^\pi(c_j)$ for every $j \neq k$. Since every outgoing action from $x$ has the same reward this also implies that the action $(x, c_k)$ is the most appealing action at $x$.

For the states $y$, we must show that the most appealing action is $(y, c_k)$, where $k = \min(B \cup n+1)$. When $C = B \cup \{n+1\}$, Proposition 33 implies that $\mathrm{Val}^\pi(c_k) > \mathrm{Val}^\pi(c_j)$ for every $j \neq k$. Since every outgoing action from $y$ has the same reward this also implies that the action $(y, c_k)$ is the most appealing action at $y$. □

## G  Proof of Proposition 13

*Proof.* For every state other than $b_i$ the proof is identical to the proof that policy iteration moves from the policy $\pi_j^B$ to the policy $\pi_{j+1}^B$ for $j < 2i$. To complete the proof we must argue that the action $(b_i, a_i)$ is the most appealing action at the state $b_i$. We will first prove that every action other than $(b_i, a_i)$ cannot be switched by policy iteration at the state $b_i$. For the actions $(b_i, d_k)$ we have by Proposition 22:

$$\mathrm{Appeal}^{\pi_{2i+1}^B}(b_i, d_k) = \mathrm{Val}^{\pi_{2i+1}^B}(y) + 4n + k + 1$$
$$\leq \mathrm{Val}^{\pi_{2i+1}^B}(y) + 4n + 2i + 1$$
$$= \mathrm{Appeal}^{\pi_{2i+1}^B}(b_i, d_{2i}) = \mathrm{Val}^{\pi_{2i+1}^B}(b_i)$$

This implies that no action of the form $(b_i, d_k)$ can be switched by policy iteration. Similarly, for the action $(b_i, y)$ we have:

$$\mathrm{Appeal}^{\pi_{2i+1}^B}(b_i, y) = \mathrm{Val}^{\pi_{2i+1}^B}(y) + 1 < \mathrm{Val}^{\pi_{2i+1}^B}(y) + 4n + 2i + 1 = \mathrm{Val}^{\pi_{2i+1}^B}(b_i)$$

For the action $(b_i, x)$, Proposition 29 gives:

$$\mathrm{Appeal}^{\pi_{2i+1}^B}(b_i, x) = \mathrm{Val}^{\pi_{2i+1}^B}(x) < \mathrm{Val}^{\pi_{2i+1}^B}(y) < \mathrm{Val}^{\pi_{2i+1}^B}(b_i)$$

Proposition 30 implies that the actions $(b_i, f_j)$ cannot be switched by policy iteration.

Now we must prove that the action $(b_i, a_i)$ can be switched by policy iteration. We begin by showing that $\mathrm{Val}^{\pi^B_{2i+1}}(g_i) > \mathrm{Val}^{\pi^B_{2i+1}}(b_i)$. Let $k = \min(B^{>i} \cup \{n+1\})$ be the smallest index in $B$ that is bigger than $i$, or the index of the sink if $i$ is the highest bit. Using Proposition 22, the fact that $i \leq n$, and Proposition 25 gives:

$$\begin{aligned}
\mathrm{Val}^{\pi^B_{2i+1}}(b_i) &\leq 6n + 1 + \mathrm{Val}^{\pi^B_{2i+1}}(y) \\
&= 6n + 1 + \sum_{j \in B^{\leq i}} (10n+4)(2^j - 2^{j-1}) + \mathrm{Val}^{\pi^B_{2i+1}}(c_k) \\
&\leq 6n + 1 + \sum_{j=1}^{i-1}(10n+4)(2^j - 2^{j-1}) + \mathrm{Val}^{\pi^B_{2i+1}}(c_k) \\
&= 6n + 1 + (10n+4)(2^{i-1} - 2^0) + \mathrm{Val}^{\pi^B_{2i+1}}(c_k) \\
&= (10n+4)2^{i-1} - 4n - 3 + \mathrm{Val}^{\pi^B_{2i+1}}(c_k)
\end{aligned}$$

The value of the state $g_i$ is:

$$\mathrm{Val}^{\pi^B_{2i+1}}(g_i) = (10n+4)2^i + \mathrm{Val}^{\pi^B_{2i+1}}(g_i) = (10n+4)2^i - 1 + \mathrm{Val}^{\pi^B_{2i+1}}(c_k)$$

Since $(10n+4)2^i - 1 > (10n+4)2^{i-1} - 4n - 3$ for every $i$, we have that $\mathrm{Val}^{\pi^B_{2i+1}}(g_i) > \mathrm{Val}^{\pi^B_{2i+1}}(b_i)$. Now we can conclude:

$$\begin{aligned}
\mathrm{Appeal}^{\pi^B_{2i+1}}(b_i, a_i) &= (1 - \frac{2^{-n}}{10n+4}) \mathrm{Val}^{\pi^B_{2i+1}}(b_i) + \frac{2^{-n}}{10n+4} \mathrm{Val}^{\pi^B_{2i+1}}(g_i) \\
&> (1 - \frac{2^{-n}}{10n+4}) \mathrm{Val}^{\pi^B_{2i+1}}(b_i) + \frac{2^{-n}}{10n+4} \mathrm{Val}^{\pi^B_{2i+1}}(b_i) \\
&= \mathrm{Val}^{\pi^B_{2i+1}}(b_i)
\end{aligned}$$

Therefore, the action $(b_i, a_i)$ will be switched by policy iteration. □

## H   Proof of Proposition 14

*Proof.* For the state $b_i$ the proof that $a_i$ is the most appealing action is identical to the proof given for Proposition 13. For every state other than $b_i$, $c_i$, $x$, or the states $b_j$ with $j < i$, the proof that policy iteration moves from $\pi^B_{2i+2}$ to $\pi^B_{2i+3}$ is identical to the proof given for Proposition 12.

For the state $c_i$ we must show that the action $(c_i, f_i)$ is the most appealing action. Using Proposition 2, and the fact that $2^i - 2^{i-1} > 0$ for every $i$ gives:

$$\begin{aligned}
\mathrm{Appeal}^{\pi^B_{R2}}(c_i, f_i) &= (10n+4)(2^i - 2^{i-1}) + 1 + \mathrm{Val}^{\pi^B_{R2}}(r_i) \\
&> \mathrm{Val}^{\pi^B_{R2}}(r_i) = \mathrm{Appeal}^{\pi^B_{R2}}(c_i, r_i)
\end{aligned}$$

Therefore $(c_i, f_i)$ is the most appealing action at the state $c_i$.

For the state $x$ we must show that the action $(x, f_i)$ is the most appealing action. Note that $\pi_{R1}^B = \pi_{2i+2}^B$ for every state except the state $b_i$. A proof that is identical to the one given for Proposition 35 can be used to conclude that if $k = \min(B \cup \{n+1\})$ then $\text{Val}^{\pi_{R1}^B}(f_k) > \text{Val}^{\pi_{R1}^B}(f_j)$ for every $j \neq i$. Since every outgoing action from $x$ has the same reward it is therefore sufficient to argue that $\text{Val}^{\pi_{R1}^B}(f_i) > \text{Val}^{\pi_{R1}^B}(f_k)$.

If $l = \min(B^{>i} \cup \{n+1\})$, then Proposition 2 implies:

$$\text{Val}^{\pi_{R1}^B}(f_i) = (10n+4)(2^i - 2^{i-1}) - 4n - 1 + \text{Val}^{\pi_{R1}^B}(c_l)$$

Moreover, we can express the value of $f_k$ as:

$$\text{Val}^{\pi_{R1}^B}(f_k) = \sum_{j \in B^{<i}} (10n+4)(2^j - 2^{j-1}) - 4n - 1 + \text{Val}^{\pi_{R1}^B}(c_l)$$

$$\leq \sum_{j=1}^{i-1} (10n+4)(2^j - 2^{j-1}) - 4n - 1 + \text{Val}^{\pi_{R1}^B}(c_l)$$

$$= (10n+4)(2^{i-1} - 2^0) - 4n - 1 + \text{Val}^{\pi_{R1}^B}(c_l)$$

Since $(10n+4)(2^i - 2^{i-1}) > (10n+4)(2^{i-1} - 2^0)$ for every $i > 0$ we can conclude that $(x, f_i)$ is the most appealing action at the state $x$.

For the states $b_j$ with $j < i$ we must show that the action $(b_j, f_i)$ is the most appealing action. A proof that is identical to the proof given for Proposition 10 can be used to show that the action $(b_i, x)$, the action $(b_j, y)$, and the actions of the form $(b_j, d_k)$ will not be switched by policy iteration in the policy $\pi_{R1}^B$. Moreover, a proof that is identical to the proof of Proposition 30 can be used to show that the actions of the form $(b_j, f_k)$ with $k \neq i$ cannot be switched by policy iteration in the policy $\pi_{R1}^B$. To complete the proof we must therefore argue that $\text{Appeal}^{\pi_{R1}^B}(b_j, f_i) > \text{Val}^\pi(b_j)$.

We have previously derived an expression for the value of $f_i$ in terms of the state $c_l$, where $l = \min(B^{>i} \cup \{n+1\})$. We can use this to obtain:

$$\text{Appeal}^{\pi_{R1}^B}(b_j, f_i) = (10n+4)(2^i - 2^{i-1}) + \text{Val}^{\pi_{R1}^B}(c_l)$$

$$= (10n+4)2^{i-1} + \text{Val}^{\pi_{R1}^B}(c_l)$$

We can also express the value of the state $b_j$ as:

$$\text{Val}^{\pi_{R1}^B}(b_j) = (10n+4)2^j - 1 + \sum_{k \in B^{>j} \cap B^{<j}} (10n+4)(2^k - 2^{k-1}) + \text{Val}^{\pi_{R1}^B}(c_l)$$

$$\leq (10n+4)2^j - 1 + \sum_{k=j+1}^{i-1} (10n+4)(2^k - 2^{k-1}) + \text{Val}^{\pi_{R1}^B}(c_l)$$

$$= (10n+4)2^j - 1 + (10n+4)(2^{i-1} - 2^j) + \text{Val}^{\pi_{R1}^B}(c_l)$$

$$= (10n+4)2^{i-1} - 1 + \text{Val}^{\pi_{R1}^B}(c_l)$$

Since $(10n + 4)2^{i-1} > (10n + 4)2^{i-1} - 1$ we have that $\text{Appeal}^{\pi_{R1}^B}(b_j, f_i) > \text{Val}^{\pi_{R1}^B}(b_j)$. This implies that the action $(b_j, f_i)$ will be switched by policy iteration at every state $b_j$ with $j < i$. □

## I  Proof of Proposition 15

**Proposition 37.** *We have* $\text{Val}^{\pi_{R2}^B}(y) + 6n + 1 < \text{Val}^{\pi_{R2}^B}(x)$.

*Proof.* Let $l = \min(B \cup \{n+1\})$. We first consider the case where $l < i$. It is not difficult to see that if $l < i$ then $l = 1$, since $i$ is the smallest index that is not contained in $B$. In this case we can express the value of $y$ in terms of the value of the state $f_i$ as:

$$\begin{aligned}\text{Val}^{\pi_{R2}^B}(y) &= -(10n+4)2^{l-1} + 4n + 2 + \text{Val}^{\pi_{R2}^B}(f_i) \\ &= -(10n+4) + 4n + 2 + \text{Val}^{\pi_{R2}^B}(f_i) \\ &= -(10n+4) + 4n + 2 + \text{Val}^{\pi_{R2}^B}(f_i) \\ &= -6n - 2 + \text{Val}^{\pi_{R2}^B}(f_i)\end{aligned}$$

Moreover, we can express the value of $x$ as:

$$\text{Val}^{\pi_{R2}^B}(x) = \text{Val}^{\pi_{R2}^B}(f_i)$$

Therefore, we have $\text{Val}^{\pi_{R2}^B}(y) + 6n + 1 < \text{Val}^{\pi_{R2}^B}(x)$.

The second case that we must consider is when $l > i$, which occurs only when $i = 1$. In this case we can express the value of $y$ in terms of the value of $c_l$ as:

$$\text{Val}^{\pi_{R2}^B}(y) = \text{Val}^{\pi_{R2}^B}(c_l)$$

Similarly, we can express the value of $x$ in terms of the value of $c_l$. Our derivation uses the fact that $i = 1$.

$$\begin{aligned}\text{Val}^{\pi_{R2}^B}(x) &= -(10n+4)2^{i-1} - 4n + (10n+4)2^i - 1 + \text{Val}^{\pi_{R2}^B}(c_l) \\ &= (10n+4) - 4n - 1 + \text{Val}^{\pi_{R2}^B}(c_l) \\ &= 6n + 3 + \text{Val}^{\pi_{R2}^B}(c_l)\end{aligned}$$

Once again it is clear that $\text{Val}^{\pi_{R2}^B}(y) + 6n + 1 < \text{Val}^{\pi_{R2}^B}(x)$. □

*Proof (of Proposition 15).* For the states in the set $\{c_j, r_j \;:\; j \geq i\} \cup \{b_j \;:\; j \in B \text{ and } j \geq i\} \cup \{x\}$ the proof that policy iteration does not switch away from the action chosen by $\pi_0^{B'}$ is identical to the proof given for Proposition 12. For the states $r_j$ with $j < i$, the proof that $(r_j, c_i)$ is the most appealing action at $r_j$ is identical to the proof given for Proposition 12. The proof that the states $b_j$ with $j \leq i$ do not switch away from the action $(b_j, f_i)$ is very similar to the proof given for Proposition 14 that $(b_j, f_i)$ was the most appealing action in the policy $\pi_{R1}^B$.

For the state $c_j$ with $j < i$ we must show that policy iteration does not switch away from the action $(c_j, f_j)$. The first case that we consider is when there is some other index $l \in B$ in the range $j < l < i$. In this case we have:

$$\text{Appeal}^{\pi_{R2}^B}(c_j, r_j) = -(10n+4)2^{l-1} + 4n + 1 + \text{Val}^{\pi_{R2}^B}(f_i)$$
$$< -(10n+4)2^{j-1} + 4n + 2 + \text{Val}^{\pi_{R2}^B}(f_i)$$
$$= \text{Appeal}^{\pi_{R2}^B}(c_j, f_j)$$

The other case that must be considered is when $j$ is the largest index in $B$ that is smaller than $i$. If $l = \min(B \cup \{n+1\})$ then we have:

$$\text{Appeal}^{\pi_{R2}^B}(c_j, r_j) = \text{Val}^{\pi_{R2}^B}(c_l) - 1$$
$$< \text{Val}^{\pi_{R2}^B}(c_l) + (10n+4)(2^i - 2^{i-1} - 2^{j-1}) + 4n + 1$$
$$= \text{Appeal}^{\pi_{R2}^B}(c_j, f_j)$$

For the state $y$ we must show that the most appealing action is $(y, c_i)$. For the actions of the form $(y, c_j)$ with $j > i$ we can use the same argument that was used in the proof of Proposition 12 to argue that $\text{Appeal}^{\pi_{R2}^B}(y, c_i) > \text{Appeal}^{\pi_{R2}^B}(y, c_j)$. For the actions $(y, c_j)$ with $j < i$ we have $\pi_{R2}^B(c_j) = f_j$ and $\pi_{R2}^B(b_j) = f_i$. Therefore, we can express the appeal of $(y, c_j)$ as:

$$\text{Appeal}^{\pi_{R2}^B}(y, c_j) = -(10n+4)2^{j-1} + 4n + 2 + \text{Val}^{\pi_{R2}^B}(f_i)$$

Since $\pi_{R2}^B(c_i) = f_i$ we can also express the appeal of the action $(y, c_i)$ as:

$$\text{Appeal}^{\pi_{R2}^B}(y, c_i) = 4n + 1 + \text{Val}^{\pi_{R2}^B}(f_i)$$

Since $-(10n+4)2^{j-1} + 1 < 0$ for every $j > 0$ we have that the action $(y, c_i)$ is the most appealing action at $y$.

For the states $d_k$ we must show that the most appealing action is $(d_k, x)$. Proposition 37, and the fact that $r(d_k, x) = r(d_k, y)$ for every $k$, imply:

$$\text{Appeal}^{\pi_{R2}^B}(y) = r(d_k, y) + \text{Val}^{\pi_{R2}^B}(y) < r(d_k, y) + \text{Val}^{\pi_{R2}^B}(x) = \text{Appeal}^{\pi_{R2}^B}(x)$$

Every state $d_k$ with $k \geq 1$ has an additional action $(d_k, d_{k-1})$, for which we consider two cases. When $1 \leq k \leq 2i + 4$ we have by Proposition 19 and Proposition 37 give:

$$\text{Appeal}^{\pi_{R2}^B}(d_k, d_{k-1}) = 4n - k + 1 + \text{Val}^{\pi_{R2}^B}(y) < \text{Val}^{\pi_{R2}^B}(x) = \text{Appeal}^{\pi_{R2}^B}(d_k, x)$$

In the case where $k > 2i + 4$, Proposition 20 and Proposition 37 imply:

$$\text{Appeal}^{\pi_{R2}^B}(d_k, d_{k-1}) = \text{Val}^{\pi_{R2}^B}(y) - 1 < \text{Val}^{\pi_{R2}^B}(x) = \text{Appeal}^{\pi_{R2}^B}(d_k, d_{k-1})$$

We have therefore shown that the action $(d_k, x)$ is the most appealing action at the state $d_k$.

Finally, we must show that the action $(b_j, x)$ is the most appealing action at every state $b_j$ with $j \notin B \setminus \{i\}$. The first case that we consider is when $j > i+1$. In this case a proof that is identical to the proof given for Proposition 8 can be used to show that the action $(b_j, d_{2i+4})$ is more appealing than every action other than $(b_j, x)$. We must therefore argue that $\mathrm{Appeal}^{\pi_{R2}^B}(b_j, x) > \mathrm{Appeal}^{\pi_{R2}^B}(b_j, d_{2i+4})$. Using Proposition 22 and Proposition 37 gives:

$$\mathrm{Appeal}^{\pi_{R2}^B}(b_j, d_{2i+4}) \leq \mathrm{Val}^{\pi_{R2}^B}(y) + 6n + 1 < \mathrm{Val}^{\pi_{R2}^B}(x) = \mathrm{Appeal}^{\pi_{R2}^B}(b_j, x)$$

For the case where $j = i+1$ there is no action $(b_j, d_{2i+4})$. Once the techniques used in the proof of Proposition 8 can be used to show that $(b_j, a_j)$ is more appealing then every action other than $(b_j, x)$. We must argue that $\mathrm{Appeal}^{\pi_{R2}^B}(b_j, x) > \mathrm{Appeal}^{\pi_{R2}^B}(b_j, a_j)$. Using Proposition 4, Proposition 22, and Proposition 37 gives:

$$\mathrm{Appeal}^{\pi_{R2}^B}(b_j, a_j) < \mathrm{Val}^{\pi_{R2}^B}(b_j) + 1$$
$$\leq \mathrm{Val}^{\pi_{R2}^B}(y) + 6n + 2$$
$$\leq \mathrm{Val}^{\pi_{R2}^B}(x) = \mathrm{Appeal}(b_j, x)$$

Therefore, policy iteration will switch the action $(b_j, x)$ at every state $b_j$ with $j \notin B$. $\square$

## J  Proof of Proposition 16

*Proof.* For the states in the set $\{c_j : j \geq i\} \cup \{b_j : j \in B \text{ and } j \geq i\} \cup \{r_j : 1 \leq j \leq n\} \cup \{x, y\}$ the proof that policy iteration does not switch away from the action chosen by $\pi_0^{B'}$ is very similar to the proof given for Proposition 12. The proof that the action $(c_j, r_j)$ is the most appealing action at the states $c_j$ with $j < i$ is very similar to the proof given for Proposition 36

We must show that $(d_k, y)$ is the most appealing action at every state $d_k$. For the action $(d_k, d_{k-1})$ we have:

$$\mathrm{Appeal}^{\pi_{R3}^B}(d_k, d_{k-1}) = \mathrm{Val}^{\pi_{R3}^B}(x) - 1 < \mathrm{Val}^{\pi_{R3}^B}(x) = \mathrm{Appeal}^{\pi_{R3}^B}(d_k, x)$$

For the action $(d_k, y)$, the fact that $\mathrm{r}(d_k, y) = \mathrm{r}(d_k, x)$ implies:

$$\mathrm{Appeal}^{\pi_{R3}^B}(d_k, x) = \mathrm{r}(d_k, x) + \mathrm{Val}(f_i)$$
$$< \mathrm{r}(d_k, y) + 4n + 1 + \mathrm{Val}(f_i) = \mathrm{Appeal}^{\pi_{R3}^B}(d_k, y)$$

Therefore, the action $(d_k, y)$ is the most appealing action at the state $d_k$.

For the states $b_j$ with $j \notin B'$ we must show that $(b_j, y)$ is the most appealing action. We first consider the case where $j > i$. For the actions $(b_j, d_k)$ we have:

$$\mathrm{Appeal}^{\pi_{R3}^B}(b_j, d_k) \leq 4n + \mathrm{Val}^{\pi_{R3}^B}(f_i) < 4n + 2 + \mathrm{Val}^{\pi_{R3}^B}(f_i) = \mathrm{Appeal}^{\pi_{R3}^B}(b_j, y)$$

For the action $(b_j, x)$ we have:

$$\text{Appeal}^{\pi_{R3}^B}(b_j, x) = \text{Val}^{\pi_{R3}^B}(f_i) < \text{Val}^{\pi_{R3}^B}(f_i) + 4n + 2 = \text{Appeal}^{\pi_{R3}^B}(b_j, y)$$

A proof that is similar to the proof of Proposition 30 can be used to show that the actions of the form $(b_j, f_k)$ will not be switched by policy iteration. Finally, for the action $(b_j, a_j)$ we have by Proposition 4:

$$\begin{aligned}\text{Appeal}^{\pi_{R3}^B}(b_j, a_j) &< \text{Val}^{\pi_{R3}^B}(b_j) + 1 \\ &= \text{Val}^{\pi_{R3}^B}(f_i) + 1 \\ &< \text{Val}^{\pi_{R3}^B}(f_i) + 4n + 2 = \text{Appeal}^{\pi_{R3}^B}(b_j, y)\end{aligned}$$

Therefore $(b_j, y)$ is the most appealing action at the states $b_j$ with $j \notin B'$ and $j > i$.

Finally, we consider the states $b_j$ where $j \notin B'$ and $j < i$. The proof that the actions $(b_j, d_k)$ and the action $(b_j, x)$ are less appealing than the action $(b_j, y)$ is identical to the proof that was given for the states $b_j$ with $j \notin B'$ and $j > i$. For the actions $(b_j, f_k)$ with $k > i$ a proof that is similar to the proof given for Proposition 30 can be used to show that these actions will not be switched by policy iteration. For the action $(b_j, f_i)$ we have:

$$\text{Appeal}^{\pi_{R3}^B}(b_j, f_i) = 4n + 1 + \text{Val}^{\pi_{R3}^B}(f_i) < 4n + 2 + \text{Val}^{\pi_{R3}^B}(f_i) = \text{Appeal}^{\pi_{R3}^B}(b_j, y)$$

For the action $(b_j, a_j)$ we apply Proposition 4 to give:

$$\text{Appeal}^{\pi_{R3}^B}(b_j, a_j) < \text{Val}^{\pi_{R3}^B}(b_j) + 1 = \text{Val}^{\pi_{R3}^B}(f_i) + 4n + 2 = \text{Appeal}^{\pi_{R3}^B}(b_j, y)$$

Therefore $(b_j, y)$ is the most appealing action at the states $b_j$ with $j \notin B'$ and $j < i$. □

## K  Proof of Theorem 18

*Proof (of Theorem 18).* It can easily be verified that, for every policy $\pi$ that total reward criterion policy iteration algorithm considers we have, for every state $s$.

$$\mathbb{E}_s^\pi \left\{ \liminf_{N \to \infty} \frac{1}{N} \sum_{i=0}^{N} \text{r}(s_i, s_{i+1}) \right\} = 0$$

This implies that $G^\pi(s) = 0$ for every such policy. When $G^\pi(S) = 0$ the bias equation given in (3) becomes:

$$B^\pi(s) = \max_{a \in M_s} (\text{r}(s, a) + \sum_{s' \in S} p(s'|s, a) \cdot B^\pi(s'))$$

This is identical to the equation (1). Therefore, we have $\text{Val}^\pi(s) = B^\pi(S)$ for every policy $\pi$.

Policy iteration decides whether an action is switchable by lexicographically comparing the gain and bias. Since $G^\pi(s) = G^\pi(s')$ for every pair of state $s$ and $s'$, and every policy $\pi$, we have that $\sum_{s' \in S} p(s'|s,a) G^\pi(s') = G^\pi(s)$ for every policy $\pi$, every state $s$, and every action $a \in A_s$. Therefore, every decision on whether an action is switchable is always made using the bias equation. This implies that policy iteration for the average-reward criterion will behave in exactly the same way as policy iteration for the total reward criterion. □

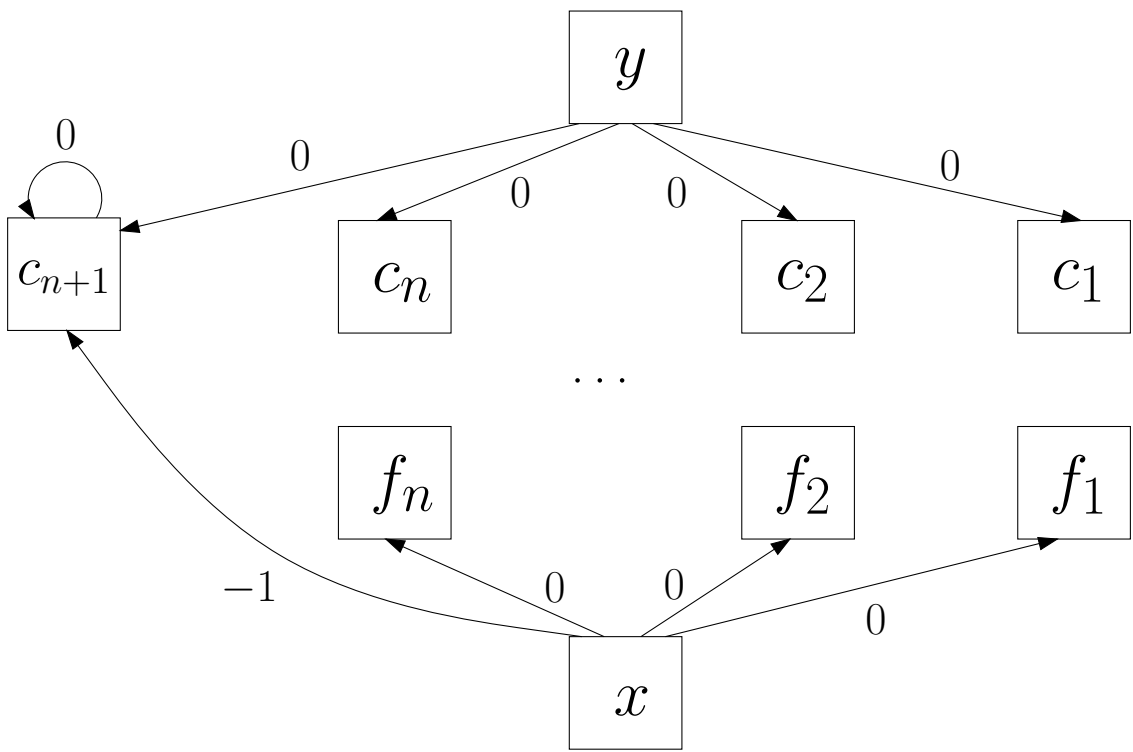